\documentclass[later,twocolumn,11pt]{article}
\usepackage[spanish]{babel}
\usepackage{ae}
\usepackage[latin1]{inputenc} 
\usepackage[T1]{fontenc} 
\usepackage{layout}
\usepackage{array}
\usepackage{graphicx}
\usepackage{amsmath}
\usepackage{multicol}
\usepackage{rotating}  
\hyphenrules{nohyphenation} 
\usepackage[]{cite}   
\usepackage{anysize}  
\usepackage{endnotes} 
\usepackage{tikz}
\marginsize{1.5cm}{1.0cm}{1.0cm}{2.5cm}
\begin{document}

\renewcommand{\tablename}{Tabla}
\renewcommand{\abstractname}{}
\renewcommand{\thefootnote}{\arabic{footnote}}

\title{Origen teórico del fotón, una introducción didáctica a la dualidad onda partícula\\
        \footnotesize   \textit{(
Theoretical photon origin, a didactic introduction to the wave particle duality)}
}
\author
{ 
Paco H. Talero L. $^{1,2}$,  William J. Robayo $^{2}$ \\
$^{1}$  \small Grupo MatCIng, Universidad Central, Cl. 21 No 4-40, Bogotá, D.C. Colombia.\\ 
$^{2}$ \small Universidad Distrital Francisco José de Caldas, Carrera 3 No. 26 A 40, Bogotá, Colombia.
}
\date{}

\twocolumn
[
\begin{@twocolumnfalse}

\maketitle 
\begin{abstract}

En cursos de física moderna es común abordar el concepto de fotón desde una perspectiva einsteniana basada en el efecto fotoeléctrico, tal formulación cuantifica
la energía del campo electromagnético, pero no profundiza en la idea de fotón como partícula.  Este trabajo mostró que es posible entender el fotón como  partícula 
al formularlo desde un punto de vista teórico sin acudir al efecto fotoeléctrico. Para esto se usó fundamentalmente la relatividad especial, la teoría electromagnética y la 
simetría inmersa en las transformaciones de Lorentz asociadas tanto a ondas electromagnéticas  monocromáticas planas como  a  partículas con masa en reposo cero. Tal enfoque 
conduce a la dualidad onda-partícula del fotón y permite proponer una didáctica alternativa para desarrollar de manera novedosa  un curso de física moderna.\\ 
\textbf{Palabras-clave:} Fotón, dualidad onda-partícula, relatividad.\\ 

In modern physics courses the idea of photon has been teaching  through from the einsteinian formulation based on the photoelectric effect.  Einstein's photon
concept  allow the quantization of the electromagnetic field,  but does not dwell on the idea of photon as particle.  The objective of this paper  was to 
demonstrate that the photon has its origin essentially in  the special relativity, the electromagnetic theory and in the symmetry between the Lorentz's
transformations associated flat monochromatic electromagnetic waves and particles with zero rest mass. This approach allows to understand the wave-particle duality of photon
and allows to propose an alternative teaching focused on this duality to develop the course of modern physics.
\end{abstract}
\end{@twocolumnfalse}
]

\section{Introducción}

En diversos textos actuales de  física moderna se introduce  el  concepto de fotón a través de una  descripción experimental fundamentada en el efecto fotoeléctrico 
e inmersa en un contexto histórico \cite{Foton,Marcelo,Esb,Serw}, lo que deja a un lado la posibilidad de realizar una aproximación a la interpretación
puramente teórica del fotón y su dualidad-onda partícula.\\
    
Atendiendo la sugerencia propuesta en \cite{Marcelo,Tejeiro} se comparó las propiedades relativistas de una partícula de masa en reposo cero (PMC) con las 
de una onda electromagnética monocromática plana (OEMP), para desvelar la simetría  presente entre estas entidades físicas y  proponer  una didáctica del 
concepto de fotón y su dualidad sin usar el efecto fotoeléctrico en su formulación.\\
    
Esta manera alternativa de formular la idea de fotón es en esencia un ejercicio  mental  que tiene como prerrequisito un hipotético escenario donde  la mecánica newtoniana, la
teoría electromagnética y  la relatividad especial han sido desarrolladas; pero, ningún descubrimiento pre-cuántica se ha realizado. Aunque  el escenario no corresponde 
al desarrollo histórico se espera sea fértil desde el punto de vista didáctico, pues resulta cómodo para el instructor de física moderna presentar la formulación alternativa
de fotón al comenzar su curso con la relatividad especial, como tradicionalmente se propone en los libros de texto \cite{Esb,Serw,Marcelo2}.\\
    
Este trabajo está organizado como sigue, en la sec.2 se exponen las propiedades relativistas de las PMC y se muestra con argumentos pre-relativistas que una OEMP
transporta momento y masa inercial; en la sec.3 se presenta y desvela la simetría entre las propiedades relativistas de una PMC y una OEMP, que conduce a la
dualidad onda-partícula asociada al fotón; en las sec.4 se propone una didáctica que permite introducir el concepto de fotón desde una perspectiva teórica 
basada en argumentos de simetría; en la sec.5 se realiza la discusión de los resultados y en en la sec.6 se concluye.\\
   
\section{Preliminares} 
    
En esta sección se presentan algunos conceptos teóricos a cerca de PMC y OEMP, los cuales son prerrequisito para realizar una introducción coherente del fotón desde un punto
de vista estrictamente teórico.
    
\subsection{Partículas con masa en reposo cero} \label{sec:pMo}
    
De la mecánica relativista \cite{Marcelo}, se conoce que  la energía total de una partícula está dada por
\begin{equation} \label{enerR}
     E=c\sqrt{m_o^2c^2+p^2}
\end{equation}
y que su velocidad se puede expresar como
\begin{equation} \label{velR}
   v=\frac{pc^2}{E} .
\end{equation}
En particular para una PMC se observa que esta se mueven con velocidad $c$, y por tanto todo observador inercial la percibe también con velocidad $c$,  lo cual se
verifica realizando una transformación de velocidades.\\
   
Ahora bien, si la masa en reposo de una partícula relativista es cero y dado que está obligada a moverse con $c$, podría pensarse que se requiere energía infinita
para acelerar hasta $c$; no es así, pues su inferencia no implica una aceleración, esto evidencia la naturaleza de tales partículas de moverse con $c$. De otro lado,
para una PMC  $E=mc^2$,  $m=m_o(1-u^2/c^2)^{-1/2}$ y $m_o=0$, ¿significa esto que siempre $m=0$ ? No, ya que al evaluar $m$ aparece una indeterminación del 
tipo $\frac{0}{0}$ que deja sin conocer  $m$, pero por consistencia obliga a que $m\neq 0$. De acuerdo con esto no es posible desde la relatividad especial conocer
teóricamente la masa de una PMC.

\subsection{El efecto Doppler}\label{sec:Dpp}
    
Considere dos sistemas de referencia inerciales $S$ y $\bar{S}$ que se mueven con rapidez relativa $v$ sobre el eje $x\bar{x}$ y una fuente de OEMP en reposo  respecto
al sistema $S$. Sea $\omega$ la frecuencia angular percibida por $S$ y $\bar{\omega}$ la percibida por $\bar{S}$, entonces usando la transformación de Lorentz y las 
invariancias del la fase $kx-\omega t=\bar{k}\bar{x}-\bar{\omega}\bar{t}$ y de $\omega=ck$ se encuentra
    \begin{equation} \label{frew}
      \bar{\omega}=\gamma \omega \left ( 1 \pm  \frac{v}{c} \right ),
    \end{equation}
siendo $\gamma$ es el factor de Lorentz\footnote{Donde $\gamma=(1-v^2/c^2)^{-1/2}$.} y $v$ la rapidez relativa. Se toma $-$ si la fuente y el observador se alejan 
y $+$ si se acercan \cite{Marcelo,Tejeiro}.

\subsection{$E=cp$ pre-relativista} \label{sec:Ecp}
    
Se quiere mostrar sin argumentos relativistas  que una OEMP transporta momento lineal. Para esto considere  una partícula con carga $q$ libre e inmersa en una OEMP que 
tiene su campo eléctrico $\vec{\varepsilon}$ en dirección $z$, su campo magnético $\vec{B}$ en dirección $x$ y se propaga en dirección $y$. Bajo estas condiciones la 
fuerza neta $\vec{F}$ sobre tal partícula está dada por fuerza de Lorentz 
    \begin{equation} \label{Fn}
        \vec{F}=\frac{q\epsilon_z v_z}{c}\widehat{u}_y+q\epsilon_z\left (1-\frac{v_y}{c}  \right )\widehat{u}_z.
    \end{equation}
Donde la partícula tiene componentes de velocidad $v_y$ y $v_z$ en las direcciones de $\widehat{u}_y$ y $\widehat{u}_z$ respectivamente y se ha usado la relación
 $B_x=\epsilon_z/c$ \cite{Marcelo}.\\
    
Así mismo, si $v_y \ll c$ y $v_z \ll c$  la energía perdida por la partícula debido a la radiación es despreciable frente a la que recibe de OEMP \cite{Marcelo}, y la
Ec.\ref{Fn} se convierte en
    \begin{equation} \label{Fnc}
        \vec{F}=\frac{q\epsilon_z v_z}{c}\widehat{u}_y+q\epsilon_z\widehat{u}_z.
    \end{equation}
Por otro lado, de la Ec.\ref{Fnc} se observa que la potencia $\varrho=\vec{F} \cdot \vec{v}$ transmitida por la OEMP a la partícula es 
    \begin{equation} \label{pote}
        \varrho=q\epsilon_z v_z,
    \end{equation}
pues $\frac{v_y}{c}\ll 1$.\\
    
Ahora, de acuerdo con la segunda ley de Newton y la Ec.\ref{Fnc}
    \begin{equation} \label{momen}
        \frac{d\vec{p}}{dt}=\frac{q\epsilon_z v_z}{c}\widehat{u}_y+q\epsilon_z\widehat{u}_z,
    \end{equation}
 siendo el segundo término irrelevante en la contribución del momento ya que al integrar la Ec.\ref{momen} el promedio sobre el periodo es cero. De manera que la 
 Ec.\ref{momen} toma la forma   
    \begin{equation} \label{momenF}
        \frac{d\vec{p}}{dt}=\frac{q\epsilon_z v_z}{c}\widehat{u}_y.
    \end{equation}
 Como $\varrho=\frac{dE}{dt}$, siendo $E$ es la energía transmitida a la partícula por la OEMP y comparando las Ecs.\ref{pote} y \ref{momenF} se encuentra
    \begin{equation} \label{p}
        \frac{d}{dt}\left ( c\vec{p}  \right )  =\frac{d}{dt}\left ( E\widehat{u}_y\right )
    \end{equation}
    cuya magnitud conduce finalmente a
    \begin{equation} \label{pf}
       p=\frac{E}{c},
    \end{equation}
 la cual corresponde al momento adquirido por la partícula en dirección de la propagación de la onda, es decir la onda ``empuja''  a la partícula.

\subsection{La caja desquiciada de Einstein, evidencia de que la luz transporta masa} \label{sec:desq}
Se trata de un experimento mental propuesto por Einstein que tiene por objeto demostrar que la radiación transporta masa inercial, para lo cual usa la mecánica
newtoniana y la teoría electromagnética, así como la conservación de la masa, la energía y el momento lineal \cite{French}.\\
     
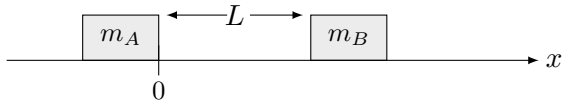
\begin{figure}[ht]
        \begin{center}
        \begin{tikzpicture}[xscale=1.0,yscale=1.0]
            \fill[gray!15](0,0) rectangle (1,0.6);
            \draw[color=black] (0,0) rectangle (1,0.6);
            \node[] at (0.5,0.3) {\small{$m_A$}};
            \fill[gray!15] (3,0) rectangle (4,0.6);
            \draw[color=black] (3,0) rectangle (4,0.6);
            \node[] at (3.5,0.3) {\small{$m_B$}};
            \draw[-latex, black] (-1cm,0cm)--(6cm,0cm);
            \draw[black] (1cm,-0.2cm)--(1cm,0.2cm);
            \node[] at (1.0,-0.4) {\small{$0$}};
            \node[] at (6.2,0.0) {$x$};
            \draw[-latex, black] (2.2cm,0.6cm)--(2.9cm,0.6cm);
            \draw[-latex, black] (1.9cm,0.6cm)--(1.1cm,0.6cm);
            \node[] at (2.0,0.6) {$L$};
         \end{tikzpicture}
        \caption{Configuración inicial del sistema, $t=0$ justo cuando la ráfaga de radiación está apunto de salir de $A$.}
        \label{luzo}
        \end{center}
    \end{figure}

Considere dos cuerpos de masas $m_A$ y $m_B$ completamente aislados y en reposo para un sistema inercial de referencia $O$, en $t=0$ del cuerpo $A$ está a punto de salir
una ráfaga de radiación hacia el cuerpo $B$, la Fig.\ref{luzo} representa esta situación.  En estas condiciones el centro de masa del sistema $x_{cm}$, es $x_{cm}=m_BL/M_T$ ya
que $A$ se encuentra en el origen.\\
     
\begin{figure}[ht]
        \begin{center}
        \begin{tikzpicture}[xscale=1.0,yscale=1.0]
            \fill[gray!15](-0.5cm,0cm) rectangle (0.5,0.6);
            \draw[color=black] (-0.5cm,0cm) rectangle (0.5,0.6);
            \node[] at (0.0,0.3) {\small{$M_A$}};
            \fill[gray!15] (3,0) rectangle (4,0.6);
            \draw[color=black] (3,0) rectangle (4,0.6);
            \node[] at (3.5,0.3) {\small{$m_B$}};
            \draw[-latex, black] (-1cm,0cm)--(6cm,0cm);
            \draw[black] (1cm,-0.2cm)--(1cm,0.2cm);
            \node[] at (1.0,-0.4) {\small{$0$}};
            \node[] at (6.2,0.0) {$x$};
            \node[] at (2.0,0.3) {$\Rightarrow$};
         \end{tikzpicture}
        \caption{La ráfaga de radiación con momento $p=E/c$, denotada con $\Rightarrow$, ya salió del cuerpo $A$ pero aún no ha llegado al cuerpo $B$.  }
        \label{luzv}\end{center}
    \end{figure}
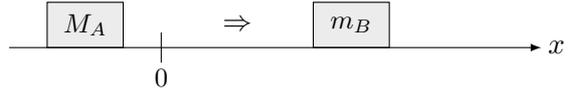

De acuerdo con Newton, todo sistema aislado  mantiene invariante su centro de masa, con esta idea debe aceptarse que la masa del cuerpo $A$ varía, de manera que justo después 
de que la radiación sale de $A$ la  nueva masa de $A$ es $M_A$, como se ilustra en la Fig.\ref{luzv}.\\
     
\begin{figure}[ht]
        \begin{center}
        \begin{tikzpicture}[xscale=1.0,yscale=1.0]
            \fill[gray!15](-1.0cm,0cm) rectangle (0.0,0.6);
            \draw[color=black] (-1.0cm,0cm) rectangle (0.0,0.6);
            \node[] at (-0.5,0.3) {\small{$M_A$}};
            \fill[gray!15] (4,0) rectangle (5,0.6);
            \draw[color=black] (4,0) rectangle (5,0.6);
            \node[] at (4.5,0.3) {\small{$M_B$}};
            \draw[-latex, black] (-1cm,0cm)--(6cm,0cm);
            \draw[black] (1cm,-0.2cm)--(1cm,0.2cm);
            \node[] at (1.0,-0.4) {\small{$0$}};
            \node[] at (0.0,-0.4) {\small{$x_A$}};
            \node[] at (4.0,-0.4) {\small{$x_B$}};
            \draw[black] (4.0cm,-0.2cm)--(4.0cm,0.0cm);
            \draw[black] (0.0cm,-0.2cm)--(0.0cm,0.0cm);
            \node[] at (6.2,0.0) {$x$};
         \end{tikzpicture}
        \caption{Configuración final del sistema, cuando ha transcurrido un tiempo $t>L/c$ y la ráfaga de radiación ha sido absorbida por el cuerpo $B$.}
        \label{luzf}
        \end{center}
    \end{figure}
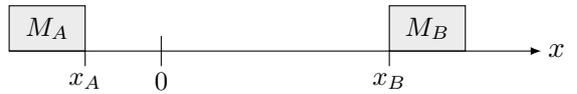

De otro lado, en la configuración inicial Fig.\ref{luzo} el momento del sistema es cero, $p=0$, y por la validez general de este principio debe permanecer así, de manera que para 
el cuerpo $A$ esto exige que $v_A=-E/M_Ac$. Así mismo, cuando la radiación alcanza el cuerpo $B$ después de un tiempo $L/c$ la conservación del momento exige que $v_B=E/M_Bc$, como 
muestra la Fig.\ref{luzf}.\\
     
De acuerdo con lo anterior, las posiciones de los bloques son
    \begin{equation} \label{cma}
     x_{A} =-\frac{E}{M_Ac}t,
    \end{equation}
y
    \begin{equation} \label{cmb}
     x_{B}=L+\frac{E}{M_Bc}\left (  t-\frac{L}{c}\right ).
    \end{equation}
Dado que el centro de masa y la masa total del sistema $M_T$ deben conservarse se tiene 
    \begin{equation} \label{cmfin}
     m_BL=M_Ax_A+M_Bx_B
    \end{equation}
y al reemplazar aquí la Ec.\ref{cma} y la Ec.\ref{cmb} se encuentra
    \begin{equation} \label{cmfin}
     -\Delta m_A=\frac{E}{c^2}=\Delta m_B.
    \end{equation}
Lo que trae como consecuencia que  una OEMP transporta masa inercial, dada por $E/c^2$.
    
\section{Fotón: relatividad y simetría}\label{sec:photon}
 
De la relatividad especial se encuentra que las transformaciones lorentzianas  de $\omega$ y $k$ para una OEMP y de $E$ y $p$ para una PMC junto con sus transformaciones inversas
tienen  forma similar, como muestra la Tab.\ref{sime}.\\ 
    
    \begin{table}[htp]
        \centering
        \begin{tabular}{c|c c}  
        OEMP &$\bar{\omega}=\gamma \left (\omega-vk \right )$ & $\bar{k}=\gamma \left (k-v\omega/c^2 \right )$ \\
        PMC& $\bar{E}=\gamma \left (E-vp \right )$ & $\bar{p}=\gamma \left (p-vE/c^2\right )$ \\ 
            &$\bar{\omega} \leftrightarrow \bar{E}$ & $ \bar{k}\leftrightarrow \bar{p}$ \\
        OEMP &$\omega=\gamma \left (\bar{\omega}+v\bar{k} \right )$ & $k=\gamma \left (\bar{k}+v\bar{\omega}/c^2 \right )$ \\
        PMC& $E=\gamma \left (\bar{E}+v\bar{p} \right )$ & $p=\gamma \left (\bar{p}+v\bar{E}/c^2\right )$ \\ 
             &$\omega \leftrightarrow E$ & $ k\leftrightarrow p$ \\ 
        \end{tabular}
        \caption{Comparación entre la transformación de magnitudes asociadas a PMC y OEMP y sus transformaciones inversas, siendo $\gamma$ es el factor de Lorentz.}
        \label{sime}
    \end{table}
Por otra parte, de la Ec.\ref{enerR} y por lo expuesto en la subs.\ref{sec:Dpp} se tienen las relaciones invariantes $E=cp$ entre energía y momento para una PMC y $\omega=ck$ para 
una OEMP, combinandolas y eliminando $c$ se encuentra
    \begin{equation} \label{ek}
       \frac{E}{\omega}=\frac{p}{k}.
    \end{equation}
Así mismo, con las ecuaciones de transformación de la Tab.\ref{sime} y la Ec.\ref{ek} se observa que $E/\omega$ y $p/k$ son invariantes relativistas.\\
    
Por consiguiente de la Tab.\ref{sime} se puede observar que las cuatro ecuaciones de transformación lorentziana y sus cuatro inversas se sintetizan en una sola con la forma
     \begin{equation} \label{uni}
      \bar{\Lambda}=\gamma  \left (\Lambda - v \Delta \right ),  \Lambda=\gamma  \left (\bar{\Lambda} + v \bar{\Delta} \right )
     \end{equation}
donde $\Lambda \leftarrow \left \{E,\omega,k,p \right \}$,  $\bar{\Lambda} \leftarrow \left \{\bar{E},\bar{\omega},\bar{k},\bar{p} \right \}$  y tanto $\Delta$ como $\bar{\Delta}$ son función de $\Lambda$ y $\bar{\Lambda}$, respectivamente.\\
   
   En otras palabras, las Ecs.\ref{uni} son invariantes a la sustitución de cada una de las ecuaciones de la Tab.\ref{sime} si se 
   especifican $\Delta(\Lambda)$ y $\bar{\Delta}(\bar{\Lambda})$, lo que se puede hacer por inspección, ver Tab.\ref{fun}. En consecuencia se tiene una 
   simetría\footnote{La definición de simetría,  es tomada  de \cite{Fey} quién afirma ``...\textit{the substance of which is that a thing is symmetrical
   if there is something we can do to it so that after we have done it, it looks the same as it did before}.''}, pues la Ec.\ref{uni} es invariante al
   transformar tanto propiedades de las OEMP  como de las PMC. Es decir, la Ec.\ref{uni} deja entrever algún tipo de unicidad entre OEMP y PMC.\\
   
   \begin{table}[htp]
        \centering
        \begin{tabular}{c c c c c}
        $\Lambda$   &$\omega$ & $k$ &$E$ & $p$ \\   
        $\downarrow$     &$\downarrow $ & $\downarrow $ &$\downarrow $ & $\downarrow $ \\   
        $\Delta$    &$k$      & $\frac{\omega}{c^2}$ &$p$ & $\frac{E}{c^2}$ \\ 
        \end{tabular}
        \caption{Asignación de $\Delta$ para cada $\Lambda$.
         Del mismo modo se asignan $\bar{\Delta},\bar{\Lambda},\bar{E},\bar{\omega},\bar{k},\bar{p}$.
           }
        \label{fun}
    \end{table}
Está simetría sugiere que coexisten características comunes de OEMP y de PMC. De ahí que sea indispensable encontrar $\Delta(\Lambda)$ y $\bar{\Delta}(\bar{\Lambda})$ de
manera más fundamental; y para esto es necesario hacer una conjetura acerca de tal dualidad. Se toma como guía la Ec.\ref{ek} y los invariantes $E/\omega$ y $p/k$ para postular
que 
   
    \begin{equation} \label{foton}
        E=\hbar \omega 
     \end{equation}
    y
   \begin{equation} \label{foton2}
        p=\hbar k,
     \end{equation}
donde $\hbar$ es una constante a determinar por los experimentos, y que por supuesto corresponde a la constante de Planck, lo cual conecta con los resultados obtenidos
con la explicación de Einstein del efecto fotoeléctrico.\\
   
Usando las Ec.\ref{uni}, \ref{foton} y \ref{foton2} se puede observar que esta conjetura resulta invariante, es decir también se cumple 
$\bar{E}=\hbar \bar{\omega}$ y $\bar{p}=\hbar \bar{k}$, que es de esperarse si de una unificación se trata. De otro lado, este resultado permite conocer la masa del fotón que de 
acuerdo con lo anterior vendrá dado por $m=\hbar \omega/c^2$, dotando de masa a una PMC lo cual no es posible sólo desde la relatividad especial. \\
   
Con las Ec.\ref{foton} y \ref{foton2} es posible obtener todas la ecuaciones de la Tab.\ref{sime} a partir de una de ellas y por tanto obtener $\Delta(\Lambda)$ y
$\bar{\Delta}(\bar{\Lambda})$  definidos en la Tab.\ref{fun}, llegando de esta manera por un camino alternativo al concepto de fotón y a la dualidad onda-partícula entre OEMP y PMC.
\section{Propuesta didáctica: origen teórico del fotón}
  
A partir de los conceptos presentados con anterioridad se propone una estrategia didáctica que permita  al instructor desarrollar el concepto de  fotón desde un punto
de vista teórico. Tal estrategia puede fundamentarse  en las cinco accciones siguientes.
\begin{enumerate}
   \item Resaltar las propiedades de las PMC y en especial que la masa de tales partículas está íntimamente relacionada con su energía mediante $E=mc^2$, ver subs.\ref{sec:pMo}.
   \item Poner en discusión las propiedades de una OEMP, en particular destacar el efecto Doppler junto con el hecho       de que estas ondas transportan momento, energía y masa
    inercial, ver subs.\ref{sec:Dpp} y subs.\ref{sec:desq}.
   \item Reflexionar a cerca de tres aspectos relacionados con el parecido entre  PMC y  OEMP: primero, las PMC están restringidas a moverse con $c$ para todo observador inercial
    lo que también hacen las OEMP; segundo, estas partículas transportan masa $E=mc^2$ y estas ondas transportan $E=\Delta mc^2$ y tercero,  tanto las PMC  como OEMP relacionan 
    su momento y energía mediante $E=cp$, ver subs.\ref{sec:Ecp}.
   \item Llamar la atención sobre las transformaciones ilustradas en la Tab.\ref{sime} para evidenciar las correspondencias $E\rightarrow \omega$ y $p \rightarrow k$  que 
   finalmente permitirán conjeturar las relaciones $E=\hbar\omega$ y $p=\hbar k$ y sus inversos, es decir la esencia del fotón y  su dualidad onda-partícula, ver sec.\ref{sec:photon}. 
\end{enumerate} 
    
\section{Discusión}
     
Se ha presentado de manera alternativa el concepto de fotón y su dualidad onda partícula a partir de la relatividad especial, la mecánica newtoniana el electromagnetismo
y argumentos de simetría, para formular una didáctica que permita entender el concepto de fotón desde un punto de vista estrictamente teórico. Lo en la práctica escolar implica tomar una
postura fiel frente al desarrollo teórico mostrado, ignorando convenientemente el desarrollo histórico para así  apreciar las consecuencias físicas de este  planteamiento y sus
ventajas didácticas.\\ 

En relación con la instrucción tradicional esta formulación presenta al menos tres ventajas didácticas: la primera consiste en resaltar las propiedades de partícula del fotón ya
que otorga la fundamentación de la dualidad onda-partícula, al provenir de una física que describe partículas, en contraste con el planteamiento de Einstein para la explicación 
del efecto fotoeléctrico que no permite inferir de manera contundente  las características del fotón como partícula; la segunda ventaja consiste en explicar el efecto
fotoeléctrico de acuerdo con la dualidad onda partícula del fotón y enfrentarse al acertijo de la interacción luz-materia para reproducir el trabajo de Einstein, pero con
una perspectiva más amplia y la tercera, dado que uno de los ejes del desarrollo es la simetría, se puede dirigir al estudiante hacia el trabajo de Luis de Broglie para generalizar
la dualidad onda-partícula, así como abordar el efecto Compton de manera más natural a la tradicional haciendo uso de $m=\hbar \omega/c^2$. Esta perspectiva permite plantear un
hilo conductor para desarrollar gran parte de un curso de física moderna con base en la búsqueda de una naturaleza simétrica y no como una conexión de hechos históricamente 
coherentes, pero sin una estricta conexión teórica global.\\
    
Aunque la formulación del fotón presentada aquí es ventajosa respecto al tratamiento tradicional, su fundamento es restringido ya que aborda un tratamiento solo de
OEMP, situación que no siempre se verifica. Aún así el planteamiento es útil para introducir la física cuántica. 

\section{Conclusiones}
A partir de la relatividad especial, la teoría electromagnética, la mecánica newtoniana y argumentos de simetría, se formuló el concepto de fotón y se hizo explícita
su dualidad onda-partícula. La cual fue fundamentada en que $E,\omega,p,k,\bar{E},\bar{\omega},\bar{p}$ y $\bar{k}$ se transforman de la forma común
$\bar{\Lambda}=\gamma  \left (\Lambda- v \Delta \right )$ y  $\Lambda=\gamma  \left (\bar{\Lambda}+ v \Bar{\Delta} \right )$, simetría que se justifica si se postula que
la naturaleza de la luz es dual y por tanto satisface $E=\hbar\omega$, $p=\hbar k$, $\bar{E}=\hbar \bar{\omega}$ y $\bar{p}=\hbar \bar{k}$, cuya implicación inmediata es dotar
de masa al fotón a través de $m=\hbar \omega/c^2$. Esta formulación alternativa mostró una didáctica novedosa y ventajosa desde lo conceptual ya que  permite desarrollar
gran parte un curso  de física moderna guiado por la búsqueda de una naturaleza simétrica.
          
\section*{Agradecimientos}
Los autores agradecen a los profesores Giovanni Cardona y Edwin Munevar por sus oportunos comentarios al trabajo. 
 
\renewcommand{\refname}{Referencias}

\end{document}